\UseRawInputEncoding
\RequirePackage{amsmath}
\documentclass[runningheads]{llncs}
\usepackage[T1]{fontenc}
\usepackage{graphicx,verbatim}
\usepackage{amssymb}
\usepackage{booktabs}
\usepackage{placeins}

\newcommand{\figref}[1]{\figurename~\ref{#1}}
\newcommand{\tabref}[1]{\tablename~\ref{#1}}
\newcommand{\secref}[1]{Section~\ref{#1}}
\newcommand{\veqref}[1]{Equation~\eqref{#1}}

\usepackage{xcolor}

\usepackage{hyperref}

\begin{document}
\title{Spatial Feature-wise Linear Modulation (SpFiLM) for Contrast Agent-Aware Brain Parcellation}
\titlerunning{Spatial Feature-wise Linear Modulation}
\author{Pushpendra Singh\inst{1}\orcidID{0009-0004-8956-1732} \and
Joshua R. Astley\inst{1}\orcidID{0000-0002-6552-5436} \and
Roman Rodionov\inst{2} \and
John Duncan\inst{2} \and
Tom Vercauteren\inst{1}\orcidID{0000-0003-1794-0456} \and
Rachel Sparks\inst{1}\orcidID{0000-0003-1553-7903}}
\authorrunning{P. Singh et al.}
\institute{School of Biomedical Engineering and Imaging Sciences, King's College London, London, UK \and
Department of Epilepsy, Queen Square Institute of Neurology, University College London, London, UK}

\maketitle              % typeset the header of the contribution
\begin{abstract}
Most automated brain parcellation tools are developed and validated on T1-weighted
(T1w) MRI. Yet, some clinical workflows for which parcellation is relevant only use contrast-enhanced T1w (T1ce) MRI, on which T1w-trained models are less accurate. We present a unified network that parcellates both pre- and post-contrast agent T1w MRI reliably, trained on a combination of the two with conditioning that spatially modulates its response differently for each. Feature-wise Linear Modulation (FiLM) is a known approach for input-based modulation in networks. It applies a per-channel scale and shift uniformly across the input.
However, the appearance change between pre- and post-contrast varies locally across the brain, making FiLM suboptimal for our use case. 
In this work, we introduce Spatial FiLM (SpFiLM), a conditioning layer whose modulation varies spatially, assembling a voxel-wise scale and shift from image-derived spatial patterns.
%SpFiLM generalizes channel-wise FiLM, which emerges as a special case when the conditioning is spatially uniform. 
Using a cohort of $134$ patients with paired T1w and T1ce MRI parcellated into 106 classes, 
%split into training, validation and held-out test sets, 
%(Mindboggle-101 DKT cortical protocol and FreeSurfer aseg subcortical segmentation), 
the addition of SpFiLM layers in a UNet increased the mean Dice on the test set of 25 patients from $80.2\%$ to $84.1\%$,
%reaches $0.841$ mean Dice, $0.039$ above the unconditioned baseline (U-Net, $0.802$).
a $4.9\%$ relative improvement. 
Adding SpFiLM layers led to the best performance on both pre- and post-contrast MRI, even when controlling for network parameter counts. 
%Ablations attribute the gains to conditioning the spatial modulation on contrast, and not to increased parameter count, spatial modulation without conditioning, or the choice of contrast encoding.

% Authors must provide keywords and are not allowed to remove this Keyword section.

\end{abstract}

\section{Introduction}
\label{sec:intro}

Brain parcellation assigns every voxel in the brain a label corresponding to a specific anatomic structure. It supports downstream clinical measurements such as region volumes and cortical thickness, both established biomarkers for neurodegenerative disease~\cite{Marek2025}, and helps define targets for radiotherapy planning~\cite{Bibault2023} and stereotactic intracerebral procedures \cite{Vakharia2019}. FreeSurfer~\cite{Fischl2012}, FMRIB Software Library (FSL)~\cite{Jenkinson2012}, Advanced Normalization Tools (ANTs)~\cite{Tustison2014}, and deep learning-based FastSurfer~\cite{Henschel2020} are widely used brain parcellation tools designed and validated on T1-weighted (T1w) MRI.

Contrast-enhanced T1w (T1ce)  MRI is acquired after administration of the contrast-agent gadolinium. As gadolinium does not cross the blood-brain barrier, structures where this barrier is lacking or disrupted show increased uptake (the intravascular compartment, choroid plexus, pituitary gland, and tumours), making T1ce useful for delineating blood vessels, tumour margins, and active lesions~\cite{Smirniotopoulos2007}.
Brain parcellation tools trained on T1w MRI have systematically different morphometric estimates between paired T1w and T1ce MRI scans~\cite{Lie2022}.

Several methods adapt neural networks to images of different appearance. Domain adaptation retrains a network to a new target domain~\cite{Rebsamen2022}.
Pre- to post-contrast domain adaptation have been proposed using distribution matching~\cite{Bermudez2020} or adversarial and paired-consistency training~\cite{OrbesArteaga2019}. 
However, domain adaptation techniques require a dedicated post-training adaptation stage to adjust network weights using target-domain data.
A single network that is reliable for both pre- and post-contrast images would remove the need to adjust network weights, or acquire and register an additional scan. This would simplify the training process, since one neural network could be applied to images regardless of contrast agent usage. A unified contrast agent-aware model could reuse common features of shared anatomy while learning contrast agent-specific differences, making more efficient use of limited training data.

Domain-randomisation methods can increase network generalisation, potentially eliminating performance differences between images of different domains, by training on synthetic images of randomised appearance so the network becomes agnostic to acquisition~\cite{Billot2023}. Such invariance suits arbitrary or unknown acquisitions, but discards usable information, since it cannot exploit systematic, anatomically localised differences between pre- and post-contrast agent MRI scans. 
Knowledge distillation can transfer shared anatomical knowledge across modalities of the same anatomy~\cite{Zhu2025}, typically assuming paired datasets.
Network conditioning can work on unpaired data by modulating a network response using information about the input images. Feature-wise Linear Modulation (FiLM)~\cite{Perez2018} is one established conditioning method that has proven effective for MRI segmentation~\cite{Lemay2021}. FiLM applies one scale and shift parameter to each feature channel, treating image differences as a global adjustment. However, the appearance differences between T1w and T1ce MRI are not uniform.

\begin{figure}[t!]
    \centering
    \includegraphics[width=1\linewidth]{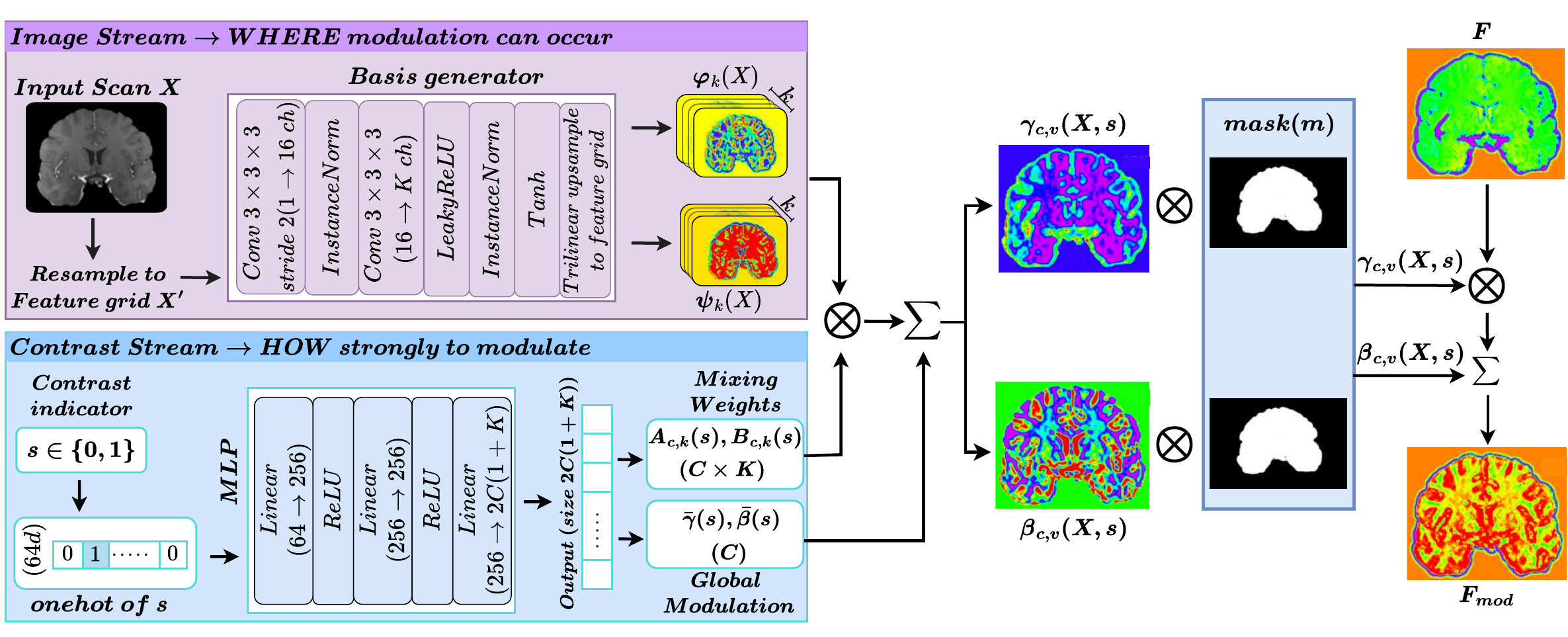}
       \caption{Overview of Spatial FiLM feature modulation block. 
       Spatial bases $\boldsymbol\varphi(X)$ and $\boldsymbol\psi(X)$ are learned from the image.
       Contrast agent indicator $s$ generates only the recombination coefficients $(\bar\gamma,\mathbf{A},\bar\beta,\mathbf{B})$. Anatomy thus determines where modulation can occur, while the contrast agent indicator determines how strongly it is applied.}
    \label{fig:overview}
\end{figure}

We introduce SpFiLM, a conditioning layer illustrated in \figref{fig:overview}, with spatially varying modulation to account for local appearance differences between T1w and T1ce MRI. In place of the channel-wise scale and shift of FiLM, it learns spatially varying scale and shift parameters but can recover channel-wise FiLM as a special case. We show that this spatial variation consistently improves brain parcellation on T1w and T1ce MRI.

\section{Methodology}
\label{sec:method}
For a set of unpaired scans, each scan $X \in \mathbb{R}^{1 \times D \times H \times W}$ had an associated contrast indicator $s\in\{0,1\}$. A convolutional neural network was trained to map $(X,s)$ to a dense label map assigning each voxel $v \in X$ to a class.
At both training and inference the network receives a single scan $X$ with its indicator $s$, never both contrasts jointly; $s$ is known from the acquisition record and is an input, not a prediction.

\subsection{Channel-wise FiLM}
\label{sec:method:film}
FiLM~\cite{Perez2018} adds a conditioning layer after each convolutional block in the encoding part of a neural network.
For a feature map $F\in\mathbb{R}^{C\times D'\times H'\times W'}$ with channels indexed by $c$,
FiLM applies the per-channel transformation:
\begin{equation}
\mathrm{FiLM}(F_{c,v}) \;=\; \bigl(1+\gamma_c(s)\bigr)\,F_{c,v} \;+\; \beta_c(s) .
\label{eq:film}
\end{equation}
The scale $\gamma_c$ and shift $\beta_c$ are produced from $s$ by a small network. 
Following Perez et al.~\cite{Perez2018}, the contrast indicator $s$ is encoded with no learned parameters: the two contrast indicators map to the first two standard basis vectors of $\mathbb{R}^{64}$.
This dimension matches the width of the learned embedding it is ablated against (\secref{sec:exp:abl}), so the two differ only in whether the mapping is learned.
A three-layer MLP with $256$, $256$, and $2C$ outputs maps the one-hot vector to one scale and one shift  value for each of the $C$ channels. Because these values carry no spatial information, every voxel in a channel applies the same scale and shift.

\subsection{Spatial FiLM}
\label{sec:method:sfilm}
SpFiLM extends \veqref{eq:film} by allowing $\gamma_c$ and $\beta_c$ to vary spatially, expressed as $\gamma_{c,v}$ and $\beta_{c,v}$ where voxels are indexed by $v$.
However, an unconstrained SpFiLM would require $\mathcal{O}(CD'H'W')$ parameters ($4.6\times10^{10}$ at our $128^3$ patch size), would tie the layer to the feature grid it was trained on, and would impose no structure on the feature modulation.
Addressing both points, we add structure to $\gamma_{c,v}$ and $\beta_{c,v}$ using a rank-$K$ factorisation to modulate features using patterns learned from the training dataset.
Feature modulation is composed of bases inferred from the image $X$, to represent spatial patterns, and contrast coefficients computed from the contrast indicator $s$, to represent patterns between contrast indicator types.

\paragraph{Image-conditioned bases.}
Each conditioning layer contains two independent basis generators, $\boldsymbol{\varphi}(X)$ and $\boldsymbol{\psi}(X)$, which produce the spatial bases for the scale and shift modulation terms, respectively. Each generator computes $K$ basis maps to capture spatial patterns present in the input image. The generators are shared across all contrast indicator types, allowing common anatomical spatial representations to be learned jointly rather than separately.

A conditioning layer modulates the feature map
$F \in \mathbb{R}^{C \times D' \times H' \times W'}$
at the encoder stage where it is inserted, with $D'$, $H'$, and $W'$ denoting the spatial dimensions at that stage. The input scan $X$ is resampled to that resolution, giving $X' \in \mathbb{R}^{1 \times D' \times H' \times W'}$.
Each basis generator then processes $X'$ using a lightweight CNN consisting of two $3 \times 3 \times 3$ convolutional blocks with instance normalisation, containing $16$ and $K$ output channels, respectively, separated by a LeakyReLU activation and followed by a final $\tanh$ activation that constrains the basis values to $[-1,1]$.

The first convolution is applied with stride $2$, so the basis maps are learned on a feature grid one resolution level coarser than $F$. They are then returned to the feature resolution by trilinear interpolation. Learning the bases on this coarser grid acts as a regulariser, encouraging smooth, spatially coherent modulation fields that operate over anatomical regions rather than introducing high-frequency, per-voxel variations. After interpolation, the resulting basis maps,
$\boldsymbol{\varphi}(X')$ and $\boldsymbol{\psi}(X') \in \mathbb{R}^{K \times D' \times H' \times W'}$,
are spatially aligned voxel-for-voxel with the feature map $F$.

\paragraph{Contrast-conditioned coefficients.}
A three-layer multi-layer perceptron (MLP) with $256$, $256$, and $2C(1{+}K)$ neurons maps the $64$-dimensional contrast indicator embedding of $s$ to spatially-invariant channel shifts ($\bar{\boldsymbol\gamma}, \bar{\boldsymbol\beta}\in\mathbb{R}^{C}$) and mixing coefficient matrices ($\mathbf{A},  \mathbf{B}\in\mathbb{R}^{C\times K}$).

\paragraph{Conditioning parameter map.}
The final conditioning parameter maps in SpFiLM are the combination of the spatially-invariant channel shifts, rank-$K$ spatial bases and contrast coefficients computed by:
\begin{equation}
\gamma_{c,v}(X, s) \;=\; \bar\gamma_c(s) \;+\; \sum_{k=1}^{K} A_{ck}(s)\,\varphi_{k,v}(X) ,
\label{eq:gamma}
\end{equation}
and similarly $\beta_{c,v}(X,s)=\bar\beta_c(s)+\sum_{k=1}^{K} B_{ck}(s)\,\psi_{k,v}(X)$.
With $K{=}0$, the sum vanishes and \veqref{eq:gamma} collapses to channel-wise
FiLM in \veqref{eq:film}.

\paragraph{Brain mask gating.}
For brain parcellation, only appearance changes within the brain are relevant for the task. Brain mask gating confines the feature modulation to anatomically meaningful voxels using a binary mask, obtained from skull-stripping (\secref{sec:method:impl}).
Skull-stripping the input does not by itself achieve this: the spatially-invariant terms $\bar\gamma_c(s)$ and $\bar\beta_c(s)$ are applied at every voxel by construction, so without gating the layer is not the identity outside the brain.
The brain mask is resampled as $m$ to each feature map $F$ with nearest-neighbour interpolation.
%, $m\in\{0,1\}^{D'\times H'\times W'}$, where $m_v$ indicates whether voxel $v$ lies inside the brain.
The final feature modulation becomes:
\begin{equation}
\mathrm{SpFiLM}(F_{c,v}) \;=\;  \bigl(1+m_v\,\gamma_{c,v}(X,s)\bigr)\,F_{c,v} \;+\; m_v\,\beta_{c,v}(X,s).
\label{eq:apply}
\end{equation}

\subsection{Implementation Details}
\label{sec:method:impl}

The backbone tested was a standard 3D U-Net~\cite{Ronneberger2015,iek2016} (five levels widths ($32$, $64$, $128$, $256$, $512$), Instance Normalization~\cite{Ulyanov2016}, LeakyReLU) with deep supervision via three auxiliary heads. All input scans were skull-stripped using HD-BET~\cite{Isensee2019-xj} and z-normalised within the resulting brain mask. Unless stated otherwise, models were trained on the combined T1w and T1ce (the \emph{both} regime), treated as unpaired so the two scan types (pre- and post contrast scans) of a patient never co-occurred in a training iteration. The $134$ patients (\secref{sec:exp:data}) were split once at the patient level into $97$/$12$/$25$ (train/val/test), giving $194$/$24$/$50$ scans with no patient shared across splits. Training used AdamW, $128^3$ patches, a deep-supervised Dice plus cross-entropy loss, and standard spatial and intensity augmentation. The SpFiLM modulation was computed in full precision with its fields $\gamma$ and $\beta$ clamped to $[-5,5]$ for numerical stability. At inference, scans were parcellated using a $128^3$ sliding window with $50\%$ overlap and Gaussian patch weighting. Full training details are available in the released code at \href{https://github.com/p-singh-kcl/spatial_film_parcellation}{https://github.com/p-singh-kcl/spatial\_film\_parcellation}.

\section{Experimental Design}
\label{sec:exp}

\subsection{Dataset}
\label{sec:exp:data}

We used a private dataset collected at the National Hospital for Neurology and Neurosurgery (NHNN), London, UK, from a prospective imaging study (REC 20/LO/0966, IRAS 278210) of $230$ adult patients with drug-resistant epilepsy.
Each patient was scanned with and without gadolinium contrast agent using a $1\,\mathrm{mm}$ isotropic MPRAGE acquisition.
Patients with anatomy distorting pathology including focal lesions, tumours, and prior surgical resection were excluded from analysis, resulting in a total of $134$ patients.

Pseudo ground truth brain parcellations were generated by multi-atlas label fusion, an approach previously used to bootstrap training labels for whole-brain segmentation~\cite{Roy2019}. The OASIS-TRT-20 Mindboggle-101 atlases~\cite{Klein2012} were propagated to each T1w MRI using NiftyReg~\cite{Modat2010}, and scan-specific labels were computed via geodesic information flow (GIF) label fusion~\cite{Cardoso2015}.
Since the atlases are T1w, gadolinium enhancement precludes fusion on T1ce: each T1ce was rigidly registered to its T1w and labels propagated by nearest-neighbour interpolation, leaving sub-voxel registration error as the dominant source of label noise.

\subsection{Model Comparisons}
\label{sec:exp:baselines}

We compared SpFiLM against four controlled baselines sharing the same U-Net backbone to isolate individual design choices.
\textbf{U-Net}, the backbone without conditioning.
\textbf{Wide U-Net} adds convolution kernels to match SpFiLM's parameter count.
\textbf{Channel-wise FiLM} applies the channel-wise conditioning in \veqref{eq:film}.
\textbf{Attention U-Net}~\cite{Oktay2018} adds multi-head attention on the decoder skip connections, applying a spatial modulation without conditioning.

\begin{table*}[tb]
\caption{Test-set Dice on 25 T1w and 25 T1ce, with params in millions.
Best in \textbf{bold}, second best \underline{underlined}.
Short names denote Ch-FiLM (channel-wise FiLM), Attn (Attention U-Net), Swin (SwinUNETR),
SpFiLM-n (narrow SpFiLM), and nnU+SpFiLM (nnU-Net+SpFiLM).
$^\dagger$Param-matched to SpFiLM $K{=}8$. $^\ddagger$Param-matched to U-Net.
nnU-Net is ResEnc-L at $300$ epochs and the conditioned models use brain-masked input.
All $\pm$ terms are across-patient standard deviations ($n{=}25$ per contrast, $n{=}50$ for All);
they exceed the between-model differences, hence the paired tests.
HD rows report the 95th-percentile Hausdorff distance (HD95, mm).
$^{*}$Per-patient mean Dice differs from SpFiLM (proposed) on both contrasts (paired $t$-test, Bonferroni $\alpha{=}0.05/16{\approx}0.003$ over 8 models $\times$ 2 contrasts, $n{=}25$).}
\label{tab:results}
\centering\footnotesize
\setlength{\tabcolsep}{3pt}
\resizebox{\linewidth}{!}{%
\begin{tabular}{l|cccc|cc|ccc}
\hline
 & \multicolumn{4}{c|}{Baselines} & \multicolumn{2}{c|}{SotA} & \multicolumn{3}{c}{Proposed} \\
\cline{2-10}
 & U-Net$^{*}$ & Wide$^{\dagger*}$ & Ch-FiLM$^{*}$ & Attn$^{*}$ & Swin$^{*}$ & nnU-Net$^{*}$ & SpFiLM-n$^{\ddagger*}$ & SpFiLM & nnU+SpFiLM \\
\hline
Params  & 22.6 & 27.6 & 23.6 & 22.7 & 35.1 & 27.5 & 22.6 & 27.7 & 30.7 \\
\hline
T1w     & 82.9 $\pm$ 2.0 & 83.0 $\pm$ 2.0 & 85.9 $\pm$ 2.0 & 85.9 $\pm$ 2.2 & 84.6 $\pm$ 2.2 & 85.7 $\pm$ 2.2 & 86.4 $\pm$ 2.0 & \underline{86.6 $\pm$ 2.0} & \textbf{87.4 $\pm$ 2.0} \\
T1ce    & 77.6 $\pm$ 3.0 & 77.7 $\pm$ 3.0 & 80.8 $\pm$ 3.1 & 81.1 $\pm$ 3.4 & 79.5 $\pm$ 3.4 & 78.9 $\pm$ 3.3 & \underline{81.4 $\pm$ 3.1} & \textbf{81.6 $\pm$ 3.0} & \textbf{81.6 $\pm$ 3.3} \\
All     & 80.2 $\pm$ 3.4 & 80.4 $\pm$ 3.4 & 83.4 $\pm$ 3.6 & 83.5 $\pm$ 3.5 & 82.1 $\pm$ 3.5 & 82.3 $\pm$ 3.3 & 83.9 $\pm$ 3.6 & \underline{84.1 $\pm$ 3.6} & \textbf{84.5 $\pm$ 3.2} \\
\hline
HD T1w  & 2.22 $\pm$ 0.50 & 2.17 $\pm$ 0.48 & 2.18 $\pm$ 0.55 & 2.14 $\pm$ 0.48 & 2.43 $\pm$ 0.65 & 2.15 $\pm$ 0.55 & \underline{2.13 $\pm$ 0.46} & 2.10 $\pm$ 0.49 & \textbf{1.93 $\pm$ 0.36} \\
HD T1ce & 2.42 $\pm$ 0.49 & 2.40 $\pm$ 0.47 & 2.43 $\pm$ 0.56 & 2.40 $\pm$ 0.49 & 2.60 $\pm$ 0.61 & 2.41 $\pm$ 0.59 & \underline{2.32 $\pm$ 0.47} & 2.39 $\pm$ 0.60 & \textbf{2.17 $\pm$ 0.52} \\
\hline
\end{tabular}}
\end{table*}

\section{Experimental Results}
\label{sec:exp:results}

%The main table additionally reports our \emph{proposed} model ($K{=}8$ with the
%frozen one-hot embedding, \secref{sec:method:embed}) and a parameter-matched
%\emph{narrow} variant whose channels are reduced so that total parameters match
%the plain baseline to within 0.1\%. The primary metric is mean foreground Dice
%over 106 classes, reported for T1w, T1ce, and their average.
%All paired comparisons against the proposed model, covering the four baselines and the narrow variant on each contrast, are Bonferroni-corrected.

\begin{figure}[tb]
    \centering
    \includegraphics[width=\linewidth]{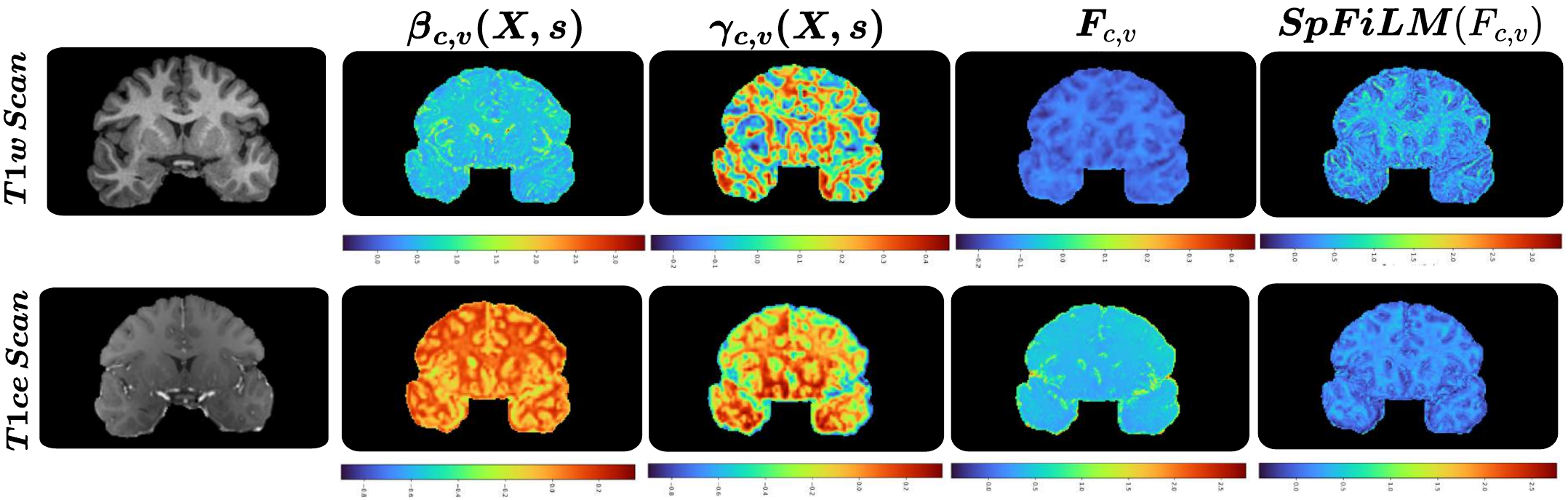}
    \caption{SpFiLM on a representative subject and feature for T1w and T1ce.
    \textbf{Modulation fields}, left to right: input scan; channel-averaged shift $\beta_{c,v}$; scale $\gamma_{c,v}$; encoder features $F_{c,v}$; modulated features $\mathrm{SpFiLM}(F_{c,v})$.}
    %\textbf{(b) Parcellation errors}: ground truth, then voxels where each model (U-Net, channel-wise FiLM, SpFiLM) disagrees with ground truth (orange); red circles mark baseline errors SpFiLM corrects (green).}
    \label{fig:mod_compare}
\end{figure}

As reported in \tabref{tab:results}, SpFiLM was the best performing model for the full dataset and for T1w and T1ce individually, reaching 84.1 mean Dice, a gain of $3.9\%$ over the U-Net baseline and $0.7\%$ over channel-wise FiLM.
Every baseline and state-of-the-art model differed significantly from SpFiLM on both contrasts (per-patient paired $t$-tests, Bonferroni-corrected; \tabref{tab:results}).
On T1w, SpFiLM lowered HD95 against the U-Net baseline, channel-wise FiLM and SwinUNETR under the paired test used for Dice; on T1ce no HD95 difference reached significance.
Parameter count alone cannot explain the increase in model performance. Wide U-Net, param count matched to SpFiLM, performed similarly to the baseline, whereas narrow SpFiLM $K{=}8$, matched to the baseline, clearly exceeded it. Attention U-Net applied spatial modulation without conditioning and only matched channel-wise FiLM.
\figref{fig:mod_compare} shows the learned fields are spatially structured and differ between T1w and T1ce.

\subsection{Comparison with the State of the Art}
\label{sec:exp:sota}
The baselines in \tabref{tab:results} share SpFiLM's backbone and bound its design choices but not its standing against established methods. We trained two widely used architectures on the same split and protocol, nnU-Net ResEnc-L~\cite{Isensee2024} and SwinUNETR~\cite{hatamizadeh2022} (\tabref{tab:results}). On a plain U-Net backbone SpFiLM already exceeded nnU-Net and the larger SwinUNETR, so its gains were not an artefact of a weak backbone. Inserting the SpFiLM layer into the nnU-Net encoder raised its mean Dice and improved T1ce most, reproducing the pooling benefit of \secref{sec:exp:abl} on a stronger backbone and confirming the conditioning is backbone-agnostic.

\subsection{Ablation Study}
\label{sec:exp:abl}
\paragraph{One-hot encoding.}
The proposed model uses a frozen one-hot encoding, in which the contrast indicator type contributes no trainable parameters. We compared this against a learned embedding, a two-layer MLP with ReLU that maps the contrast indicator type $s$ into $\mathbb{R}^{64}$ where the MLP is trained jointly with the network. As shown in \tabref{tab:abl}, the learned embedding did not improve over the frozen one-hot encoding on either contrast, model gains come from the spatial factorisation mechanism rather than the choice of contrast encoding.

\paragraph{Rank $K$.}
We evaluated the effect the number of bases, $K\in\{1,2,4,8,16\}$, had on model performance.
As shown in \tabref{tab:abl}, lower $K$ reduced SpFiLM performance.
A single basis ($K = 1$) gives one shared spatial pattern for all channels, too restrictive and slightly below channel-wise FiLM ($K = 0$). $K \geq 2$ gives enough independent spatial degrees of freedom to recover the full gain.
Because accuracy is already flat for $K \ll C$, the factorisation removes parameters the task does not use: the conditioning layers hold $5.0$M parameters at $K{=}8$ against $181$M for full-rank mixing ($K{=}C$).

\paragraph{T1w and T1ce model training.}
We compared \emph{specialist} models, each trained only on the contrast it is evaluated on, against models trained on the combined T1w and T1ce data (\emph{both}), for the unconditioned U-Net and for SpFiLM (\tabref{tab:spec}).
Pooling improved SpFiLM on T1ce by $+0.9\%$ but the U-Net by only $+0.1\%$. The \emph{both}-regime models were trained on the same data as each other, so this difference is attributable to contrast conditioning exploiting cross-contrast information rather than the larger training set, equally available to both. On T1w the gains were small for both ($+0.4\%$ and $+0.3\%$), confining the benefit to T1ce.

\begin{table}[tb]
\caption{Contrast-specialist training (each model trained only on the contrast it is evaluated on) versus joint training on \emph{both} contrasts. Mean Dice (\%) on the test set.}
\label{tab:spec}
\centering\footnotesize
\setlength{\tabcolsep}{8pt}
\renewcommand{\arraystretch}{1.2}
\begin{tabular}{l|cc|cc}
\hline
       & \multicolumn{2}{c|}{T1w} & \multicolumn{2}{c}{T1ce} \\
\cline{2-3} \cline{4-5}
Model  & Specialist & Both       & Specialist & Both        \\
\hline
U-Net  & 82.6       & \textbf{82.9} & 77.5    & \textbf{77.6} \\
SpFiLM & 86.2       & \textbf{86.6} & 80.7    & \textbf{81.6} \\
\hline
\end{tabular}
\end{table}

\begin{table}[tb]
\caption{Ablation and sensitivity analysis for SpFiLM. Rank $K$ (left) and insertion level (middle) are sensitivity sweeps, showing stable performance for $K \geq 2$ and across depth: conditioning only the shallowest level already recovers most of the improvement ($83.7$ versus $84.1$). Contrast encoding (right) is an ablation. $\dagger$ marks the proposed configuration. Mean Dice (\%) on the held-out test set.}
\label{tab:abl}
\centering\footnotesize
\setlength{\tabcolsep}{3pt}
\begin{tabular}{c | rrr | c rrr | l rrr}
\hline
$K$ & T1w & T1ce & Mean & Levels & T1w & T1ce & Mean & Encoding & T1w & T1ce & Mean \\
\hline
1  & 85.6 & 80.4 & 83.0 & Top-1 & 86.2 & 81.2 & 83.7 & One-hot$^\dagger$ & 86.6 & 81.6 & 84.1 \\
2  & 86.4 & 81.4 & 83.9 & Top-2 & 86.2 & 81.3 & 83.8 & Learned  & 86.4 & 81.5 & 84.0 \\
4  & 86.6 & 81.5 & 84.0 & Top-3 & 86.5 & 81.4 & 83.9 &          &      &      &      \\
8$^\dagger$ & 86.6 & 81.6 & 84.1 & All$^\dagger$  & 86.6 & 81.6 & 84.1 &          &      &      &      \\
16 & 86.6 & 81.5 & 84.1 &      &      &      &      &          &      &      &      \\
\hline
\end{tabular}
\end{table}

\section{Conclusion}
\label{sec:conclusion}

We presented SpFiLM, a conditioning layer that adjusts a spatially varying feature modulation to image contrast, with channel-wise FiLM recovered as the $K{=}0$ special case. Ablation studies attribute SpFiLM's improvement to its contrast-conditioned spatial modulation, not to parameter count, spatial modulation without conditioning, or the choice of contrast encoding. SpFiLM surpassed nnU-Net and SwinUNETR baselines, and improved nnU-Net further when used in the nnU-Net backbone, indicating the mechanism is backbone-agnostic.
A single network trained jointly on T1w and T1ce MRI, with no per-contrast models, was the best performing model on both contrasts and, unlike the unconditioned baseline, converted the pooled cross-contrast data into a gain on the clinically-acquired T1ce.

However, our study has limitations. The cohort is from a single centre and scanner, limiting generalisability. The reference labels are atlas-generated rather than manually annotated, so the atlas accuracy bounds the model and our findings. Because gadolinium is not given to healthy volunteers, paired pre- and post-contrast scans are available only from patients with a clinical indication, we therefore excluded patients with anatomy-distorting pathology (\secref{sec:exp:data}) so that contrast-handling errors are not confounded with pathology-induced label errors. Evaluating SpFiLM on pathological cohorts and validating it across institutions and scanners are natural next steps, as is a direct comparison against an unfactorised SpFiLM, which we defer to a journal extension. Finally, the method is not specific to brain parcellation, any segmentation task where contrast-agent uptake produces spatially varying intensity could benefit from SpFiLM.

\begin{credits}
\subsubsection{\ackname} This work was funded by the Engineering and Physical Sciences Research Council [Grant Number EP/Y035364/1] and a Hinduja PhD Scholarship and the National Institute for Health and Care Research (NIHR) Invention for Innovation Programme (NIHR208398). The views expressed are those of the author(s) and not necessarily those of the NIHR or the Department of Health and Social Care.

\subsubsection{\discintname} 
TV is co-founder and shareholder of Hypervision Surgical who have no interest in this work.
\end{credits}

\bibliographystyle{splncs04}
\bibliography{spa}

\end{document}